\documentclass[fleqn,usenatbib]{mnras}

\usepackage{newtxtext,newtxmath}

\usepackage[T1]{fontenc}

\DeclareRobustCommand{\VAN}[3]{#2}
\let\VANthebibliography\thebibliography
\def\thebibliography{\DeclareRobustCommand{\VAN}[3]{##3}\VANthebibliography}

\usepackage{graphicx}	
\usepackage{amsmath}	
\usepackage{amsmath} 
\usepackage{float} 
\usepackage{graphicx} 
\usepackage{booktabs} 
\usepackage{hyperref} 
\usepackage{geometry}
\usepackage{booktabs} 
\usepackage{ifthen}
\usepackage{lipsum}
\usepackage{color, soul} 
\usepackage{makecell}
\usepackage{subcaption}
\usepackage{xspace} 

\usepackage[normalem]{ulem} 
\usepackage[dvipsnames]{xcolor}

\newcommand{\eq}[2][]{\begin{align} #2 \end{align}}

\newcommand{\abs}[1]{\left | #1 \right |}
\newcommand{\avg}[1]{\left \langle #1 \right \rangle}
\newcommand{\diff}{\mathrm{d}}

\newcommand{\lr}[1]{\left(#1\right)}

\newcommand{\BH}{\mathrm{BH}}
\newcommand{\BBH}{\mathrm{BBH}}
\newcommand{\Sun}{\odot}
\newcommand{\nbody}{\texorpdfstring{$N$-body\xspace}{N-body\xspace}}
\newcommand{\NBODY}{\texorpdfstring{$N$-BODY\xspace}{N-BODY\xspace}}

\def\subinrm#1{\sb{\rm#1}}
{\catcode`\_=13 \global\let_=\subinrm}
\mathcode`\_="8000
\def\upsubscripts{\catcode`\_=12 } 

\upsubscripts

\newcommand{\msun}{{\rm M}_\odot}

\newcommand{\eqrefnp}[1]{equation~(\ref{#1})}
\renewcommand{\eqref}[1]{(equation~\ref{#1})}

\title[Demographics of 3-body BBHs in star clusters]{Demographics of three-body binary black holes in star clusters: implications for gravitational waves}

\author[D. Mar\'in Pina et al.]{
Daniel Mar\'in Pina\thanks{E-mail: danielmarin@icc.ub.edu}$^{1}$ and
Mark Gieles$^{1,2}$
\\
$^{1}$Departament de F\'isica Qu\`antica i Astrof\'isica, Institut de Ci\`encies del Cosmos, Universitat de Barcelona, Mart\'i i Franqu\`es 1, E-08028 Barcelona, Spain\\
$^{2}$ICREA, Pg. Llu\'is Companys 23, E08010 Barcelona, Spain
}

\date{Accepted XXX. Received YYY; in original form ZZZ}

\pubyear{2023}

\begin{document}
\label{firstpage}
\pagerange{\pageref{firstpage}--\pageref{lastpage}}
\maketitle

\begin{abstract}
To explain both the dynamics of a globular cluster and its production of gravitational waves from coalescing binary black holes, it is necessary to understand its population of dynamically-formed (or, `three-body') binaries. We provide a theoretical understanding of this population, benchmarked by direct \nbody models. We find that \nbody models of clusters on average have only one three-body binary at any given time. This is different from theoretical expectations and models of binary populations, which predict a larger number of binaries ($\sim 5$), especially for low-$N$ clusters ($\sim 100$), or in the case of two-mass models, low number of black holes. We argue that the presence of multiple binaries is suppressed by a high rate of binary-binary interactions, which efficiently ionise one of the binaries involved. These also lead to triple formation and potentially gravitational wave (GW) captures, which may provide an explanation for the recently reported high efficiency of in-cluster mergers in models of low-mass clusters ($\lesssim10^5\,\msun)$.
\end{abstract}

\begin{keywords}
stars: kinematics and dynamics -- black hole physics -- binaries: general -- gravitational waves
\end{keywords}



\section{Introduction}
The first detection of a gravitational wave (GW) signal \citep{LVK2016} was one of the key milestones for the advent of the multimessenger era of astrophysics. This has allowed the direct study of, among others, dark systems such as black hole (BH) binaries. To date, 90 GW signals from mergers of binary compact objects have been detected \citep{LVK2021a, LVK2021b} -- of which the vast majority are binary black holes (BBH) -- which requires us to tackle the question of the origin of these sources. 

Dynamical interactions in the dense cores of stellar clusters  have been put forward \citep{Zwart2000} as one of the most likely formation channels for BBH mergers \citep{Rodriguez2022, Kremer2020, Chattopadhyay2022, AntoniniGieles2020b, Kumamoto2020, Sambaran2021b, Mapelli2022}. Current GW constraints are compatible with a majority of massive mergers having formed dynamically \citep{Antonini2023}. The predictions for the production of BBH mergers in this channel hinge on the population of BBHs, which points towards the importance of fully characterising the demographics of these systems.
Because the majority of massive stars form in binaries or higher order multiples \citep{2012Sci...337..444S, 2017ApJS..230...15M}, a large fraction of BHs end up with a companion after stellar evolution (hereafter `primordial BBHs').
It has been suggested, however, that in massive clusters the binaries that result in  BBH mergers are dominated by the three-body\footnote{We note that \citet*{Tanikawa2012} showed that their formation  tends to involve more than three particles, but for historic reasons we prefer to refer to dynamically-formed binaries from single bodies as three-body binaries} BBHs \citep{Chattopadhyay2022, 2022MNRAS.517.2953T}. A possible explanation of this is due to primordial binaries typically having a smaller semi-major axis, and thus a smaller interaction cross-section, than three-body binaries \citep{Barber2023}. It is, therefore, important to study the demographics of three-body binaries to understand dynamical mergers.

 \cite{Goodman1984} showed that in steady post-collapse evolution, the number of 
 three-body binaries, $N_b$,
 depends on the total number of stars in the cluster, $N$,  as $N_b\propto N^{-1/3}$. These  models predict that a handful of binaries is expected in single-mass clusters of a few hundred  stars. However, \nbody models of such clusters find instead that there is typically only one binary present \citep{GierszHeggie1994b}. The aim of this paper is to understand the demographics of the population of three-body binaries in single-mass and two-mass star clusters, where the latter are intended to understand the behaviour of clusters with stellar-mass BHs. We do this by revisiting the model for steady post-collapse evolution and by comparing this to results from direct \nbody calculations.

The paper is organised as follows: in Section~\ref{section:model} we revisit the theory for the population of three-body binaries. In Section~\ref{section:nbody} we present the \nbody simulations, of single-mass and two-mass models. In Section~\ref{section:comparing} we compare the predictions and the \nbody models. The cause of the aforementioned discrepancy is discussed in Section~\ref{section:binBHbinBHinteractions}. In Section~\ref{sec:discussion} we discuss its implications for GW observations.

\section{The number of three-body binaries in steady post-collapse evolution}
\label{section:model}
In this section we revisit the models for steady post-collapse evolution of both single-mass \citep{Goodman1984} and two-mass \citep{BreenHeggie2013} clusters.

Throughout this paper we consider star clusters in the post-core-collapse evolutionary phase, when binary formation and hardening is ongoing and there is a balance between the energy production by binaries in the core and energy transport by two-body relaxation \citep{Henon1975}. Furthermore, we assume that this post-collapse evolution can be approximated by a steady evolution of the cluster parameters, i.e. we ignore gravothermal oscillations that occur in single-mass models with $N\gtrsim 8000$ stars \citep{BettwieserSugimoto1984, Breeden1994}
and in two-mass models where the number of BHs is $N_\BH \gtrsim 2000$ \citep{BreenHeggie2012a}.
Both of these assumptions allow us to express relations between several cluster properties, which are discussed below and summarised in Table \ref{tab:scalinglaw} and Table \ref{tab:scalinglawmultimass}.

\subsection{Single-mass clusters}
\label{ssec:sssc}

\begin{table}
\centering
\caption{Scaling laws for single-mass cluster properties in the steady post core-collapse evolutionary phase. The coefficients $\alpha_1$ and $\alpha_2$ are  found from the $N$-body simulations.}
\begin{tabular}{@{}lll@{}}
\toprule
Value                                           & Scaling law & Coefficients \\ \midrule
Velocity dispersion in the core  & $\sigma_c^2=\alpha_1\frac{2}{15}\frac{GmN}{r_h}$ & $\alpha_1 \simeq 1.4$ \\
Half-mass to core radius ratio                  & $\frac{r_h}{r_c}=\alpha_2 N^{2/3}$ & $\alpha_2 \simeq 0.13$ \\
Net creation rate of binaries                   & $\Gamma_b=0.75\frac{G^5m^5 n_c^3}{\sigma_c^9}$ & \\
Binary hardening rate                           & $\dot{z}=3.8\frac{G^2 m^2 n_c}{\sigma^3_c}$ & \\ 
Core radius & $\frac{4\pi G}{9}\rho_0r_c^2 = \sigma_c^2$ &  \\ \bottomrule
\end{tabular}
\label{tab:scalinglaw}
\end{table}

In post-collapse, we expect the cluster to be nearly isothermal within the half-mass radius $r_h$, so that the one-dimensional velocity dispersion is similar within the core and within $r_h$. For a cluster with mass $M$ and energy $E$ in virial equilibrium with a virial radius $r_v=-GM^2/(4E)$ and positions and velocities sampled from a \citet{Plummer1911} model, we have  $r_h\simeq 0.8 r_v$ and the one-dimensional velocity dispersion, $\sigma$, then obeys 
\eq{\label{singlemass:sigma2}\sigma^2 = \frac{1}{3}\langle v^2\rangle=\frac{1}{3}\frac{GM}{2r_v}\simeq \frac{2}{15}\frac{G N m}{r_h}.}
where $G$ is the gravitational constant, $m$ is the mass of the stars and $\langle v^2\rangle$ is their mean-square  velocity. The central dispersion can be expressed in terms of $\sigma$ by defining a numerical coefficient $\alpha_1$
\eq{\label{singlemass:sigma2c}\sigma_c^2=\alpha_1\sigma^2.}

Next, we define the relation between $r_h$ and the core radius, $r_c$. The corresponding equation can be obtained for the steady post collapse evolution via H\'{e}non's principle \citep{Henon1975} by noting that the energy flow through $r_h$ needs to be provided  by the core. The exact form of the $r_h/r_c$ ratio requires one to specify the energy-generating mechanism. This derivation has already been carried out in \citet[p. 265]{HeggieHut2003} and \cite{BreenHeggie2013} (although the former did not keep the numerical pre-factors) when the bulk of the energy is set to come from binary hardening via three-body processes. By assuming that the mean mass, the Coulomb logarithm ($\ln\Lambda \simeq\ln(0.11N)$,  \citealt{GierszHeggie1994a}) in the denominator of the central relaxation timescale, and the dimensionless ratio of the central potential ($|\phi_0|$) to the central velocity dispersion, $|\phi_0|/\sigma_c^2$, all are constant, they obtain
\eq{\label{singlemass:rhrc} r_h/r_c \simeq \alpha_2 N^{2/3},}
where $\alpha_2$ is a constant. If the assumptions above are relaxed, the final ratio would include a dependence on $(|\phi_0|/\sigma_c^2)/\ln\Lambda$ but, since both these quantities scale weakly with $N$, we are justified in taking $\alpha_2$ as a constant. In Section~\ref{section:comparing} we determine $\alpha_2$ and a possible additional $N$-dependence from $N$-body simulations.

We  complement the above relations with the relation for the number of stars in the cluster core, $N_c$. Starting from the  definition of the core radius \cite[p. 71]{HeggieHut2003}
\eq{\label{singlemass:sigma2king}\frac{4\pi G}{9}\rho_0r_c^2 = \sigma_c^2}
We then use the fact that, for the isothermal model and \citet{King1966} models with high concentration, the central density can be expressed in the average density within the core as $\rho_0\simeq 1.9\rho_c$, allowing us to obtain a measure of the mass contained within the core radius. Then, the number of components (singles + binaries) within the core, $N_c$, is\footnote{Unless explicitly stated otherwise, throughout this paper we assume that the binary fraction is sufficiently small such that the mean mass within the cluster is well approximated by the mass of the single stars and that $N$ is equal to the sum of the number of single stars and the number of binaries. The corrections to our results when relaxing this assumption are quantified at the end of this section.}
\eq{\label{singlemass:Nc}N_c=0.98 \frac{\alpha_1}{\alpha_2} \lr{\frac{N}{10^2}}^{1/3}}
The $\alpha_{\it i}$ parameters in this theoretical prediction are of order unity and their values and possible $N$-dependences will be determined in Section~\ref{section:comparing} by fits to \nbody models.

The binaries that we are concerned with are those that do not break up after an encounter with a typical single star of the cluster, meaning that their binding energy $x=Gm^2/(2a)$ is greater than $x_h\simeq m\sigma_c^2$ (where $a$ is the semimajor axis of the binary). We refer to these binaries with the usual nomenclature of hard binaries, as opposed to soft binaries which have $x<x_h$, and $a>a_h=Gm^2/(2 x_h)$. \citet{Heggie1975} showed that hard binaries become on average more bound (that is, they harden) after an interaction with a third star. The fractional increase of $x$ is on average $\Delta\simeq 0.4$ for resonant encounters\footnote{The fractional energy change per interaction is lower when considering all interactions, but we use the result for resonant encounters because $\Delta$ is not well-defined for non-resonant encounters \citep{Hut1993}.} \citep{Spitzer1987} and the released energy results in a velocity kick of the single star and the centre-of-mass  of the binary. Since each encounter gives the binary a momentum kick that scales with $x$, at some moment $x$ is sufficiently large such that the subsequent velocity kick of the  binary is above the escape velocity from the centre of the cluster. From conservation of energy and linear momentum and assuming equal masses, it can be shown that the upper limit of the binding energy of three-body binaries is $x_{ej}=\frac{6}{\Delta}m \abs{\phi_0}= 15 m \abs{\phi_0}$ \citep{Goodman1984}, which corresponds to a semimajor axis of $a_{ej}=Gm^2/(2 x_{ej})$
. In order to obtain the total number of binaries, we will compute the distribution of (hard) binaries per unit volume and per unit $z\equiv x/(m \sigma_c^2)$ and integrate it from $z_h=1$ to $z_{ej}=15 \abs{\phi_0}/\sigma_c^2$, where $\abs{\phi_0}/\sigma_c^2$ is a dimensionless central potential, which is  a  measure of the (instantaneous) central concentration\footnote{This quantity is reminiscent of the dimensionless central potential $W_0$ of King's model \citep{King1966}, and for $W_0\gtrsim5$ indeed $W_0\simeq|\phi_0|/\sigma_c^2$.}. This approach  results in a scaling law for the number of hard binaries \citep{Goodman1984} that we will reproduce with a detailed analysis of the numerical pre-factors.

The first model for the energy distribution of three-body binaries in post-collapse evolution was presented by \cite{Retterer1980} and was later refined by \citet{GoodmanHut1993}. We will re-derive the binary distribution using the \citeauthor{GoodmanHut1993} pre-factors, because they are based on a large set of scattering experiments by \citet{HeggieHut1993}. We begin by considering the net creation rate of binaries per unit volume \citep{GoodmanHut1993}
\eq{\label{singlemass:cbar} \Gamma_b=0.75\frac{G^5m^5 n_c^3}{\sigma_c^9},}
and the hardening rate \citep{HeggieHut1993} defined as the rate of energy generation of the binaries
\eq{\dot{z}=3.8\frac{G^2 m^2 n_c}{\sigma^3_c}.}
Since this hardening rate is independent of $z$ itself, the steady post-collapse distribution of hard binaries is uniform in $z$ between the minimum and maximum binding energy. The distribution per unit volume, $f(z)$, is 
\eq{\label{singlemass:distribution} f(z)=\frac{\Gamma_b}{\dot{z}}=0.20 \frac{G^3m^3n_c^2}{\sigma_c^6}.}
Assuming that $z_{ej}\gg z_h$ and that the relevant binary physics happens within the volume $V_c$ of the cluster core, the total number of binaries is\footnote{Compared to \citet{Goodman1984}, we use an updated value of the numerical prefactor, so their prediction for the number of binaries is roughly three times larger than ours.}
\eq{\label{singlemass:Nborig} N_b&=\int^{z_{ej}}_{z_h} f(z) V_c \diff z\simeq 3.0\frac{G^3m^3n_c^2}{\sigma_c^8} V_c \abs{\phi_0}.}

The above equation can be simplified by using the scaling laws in Table~\ref{tab:scalinglaw} to obtain\footnote{So far we have assumed that $N\simeq N_s + N_b$, with $N_s$ the number of single stars, and that the mean mass within the core, $m_c$, is equal to the mean mass of the cluster, that is, $m\simeq m_c$. If we explicitly keep the $m/m_c$ dependence in the above equations, we obtain that the result for $r_h/r_c$ (equation~\ref{singlemass:rhrc}) and $N_b$ (equation~\ref{singlemass:Nborig}) need to be multiplied by $m/m_c$ and $(m/m_c)^3$, respectively. The result for $N_c$ (equation~\ref{singlemass:Nc}) would remain unchanged. }

\eq{\label{singlemass:Nb} N_b = 2.8\frac{\alpha_2}{\alpha_1} \frac{\abs{\phi_0}}{\sigma_c^2}\left(\frac{N}{10^2}\right)^{-1/3}.}
The ratio $\abs{\phi_0}/\sigma_c^2$ is dimensionless and thus only depends on $N$. In Section~\ref{ssec:singlenb} we quantify this dependence and show that it is very weak. By neglecting it, \cite{Goodman1984} found that the scaling of the number of binaries with the number of stars in the cluster could be approximated as $N_b\propto N^{-1/3}$.

\subsection{Two-mass clusters}
\label{ssec:sstc}

\begin{table}
\centering
\caption{Scaling laws for two-mass cluster properties in the steady post core-collapse evolutionary phase. The coefficients $\alpha_{1, BBH}$ and $\alpha_{2, BBH}$ are  found from the $N$-body simulations.}
\begin{tabular}{@{}lll@{}}
\toprule
Value                                           & Scaling law & Coefficients \\ \midrule
Velocity dispersion in the core  & $\sigma_{c, \BH}^2=\alpha_{1, \BH}\frac{2}{15}
 \frac{G M_\BH}{r_{h, \BH}}$ & $\alpha_{1, \BH} \simeq 1.5$ \\
Half-mass to core radius ratio                  & $\frac{r_{h, \BH}}{r_{c, \BH}}=\alpha_{2, \BH} N_\BH^{2/3}$ & $\alpha_{2, \BH} \simeq 0.088$ \\
Net creation rate of binaries                   & $\Gamma_b=0.75\frac{G^5m_\BH^5 n_{c, \BH}^3}{\sigma_{c, \BH}^9}$ & \\
Binary hardening rate                           & $\dot{z}=3.8\frac{G^2 m_\BH^2 n_{c, \BH}}{\sigma^3_{c, \BH}}$ & \\ 
Core radius & $\frac{4 \pi G}{9} \rho_{0, \BH} r_{c, \BH}^2 = \sigma_{c, \BH}^2$ &  \\ \bottomrule
\end{tabular}
\label{tab:scalinglawmultimass}
\end{table}

In the post-collapse evolution of a cluster, the most massive stars have already ended their evolution, so the stellar population can be approximated by two distinct mass bins: that of heavy stellar-mass BHs with masses of $m_\BH\simeq20\,\msun$ and that of light stars with $m_\star\simeq0.5\,\msun$.  Due to the large difference between these bins, we will assume that all BHs have the same mass $m_\BH$ and all stars have the same mass $m_\star\ll m_\BH$. Each of these mass bins contribute a total of $M_\BH$ and $M_\star$ to the total cluster mass, respectively. We will continue referring to the heavy and light components as BHs and stars, respectively, but we note that we are considering Newtonian gravity in the point-mass limit.

To describe the two-mass cluster, we need two extra parameters with respect to the single-mass case: the mass ratio, $\mu\equiv M_\BH/M$, which sets what fraction of the cluster mass is in  BHs, and the stellar mass ratio, $q\equiv \avg{m}/m_\BH$, which sets how massive the BHs are with respect to the average mass of stars and BHs, $\avg{m}$, in the cluster. Both of these quantities are defined such that $\mu, q \in (0, 1)$. Some authors \citep[for example,][]{BreenHeggie2013} give alternative definitions of these parameters with respect to the mass of stars, $M_\star$ and $m_\star$, instead of total and average mass. Both of these definitions are interchangeable, and converge in the limit of a small number of BHs. We prefer the definition given above as it simplifies the equations that follow.

The clusters of interest will be those in which the light stars form the bulk of the cluster mass\footnote{Although this is a standard assumption, there is a sparkling interest in low-density clusters near dissolution where most of the mass is in BHs (see for example \citealt{2011ApJ...741L..12B} and the discussion about Palomar 5 in \citealt{Gieles2021}). Nevertheless, we will not focus on such systems.}, $M_\star\gg M_\BH$. This allows us to compare our results to the two-mass model of \cite{BreenHeggie2013}. Their models assume that equipartition between stars and BHs can not be achieved, which applies if \citep{1969ApJ...158L.139S}
\eq{\frac{M_\BH}{M_\star}\left (\frac{m_\BH}{m_\star}\right)^{3/2}>0.16.}

After core collapse, the BHs in a two-mass cluster behave as a denser BH subsystem embedded in a larger star cluster. The central region of the cluster will then be populated mainly by BHs, whose evolution is powered by BBHs. This energy is then transferred outwards in the BH subsystem until it is deposited in the cluster of stars surrounding the BHs. In order to obtain the equations relating the cluster properties, we will make use of the model of \cite{BreenHeggie2013}. These scaling laws, which are the analogues of equations~(\ref{singlemass:sigma2}-\ref{singlemass:Nc}) for two-mass clusters, are summarised in Table~\ref{tab:scalinglawmultimass}. The results are equivalent to what we would have found if we considered the BH sub-cluster in isolation, with the only key difference being the central potential term, which here includes the contribution of both the stars and the BHs. We reproduce the equations here, for completeness
\eq{\label{multimass:sigma2cBH} 
\sigma_{c, \BH}^2=\alpha_{1, \BH} \sigma^2_\BH = \alpha_{1, \BH}\frac{2}{15}
 \frac{G M_\BH}{r_{h, \BH}}}
\eq{\label{multimass:rhBHrcBH} \frac{r_{h, \BH}}{r_{c, \BH}}=\alpha_{2, \BH} N_\BH^{2/3}}
\eq{\label{multimass:sigma2kingBH} \frac{4 \pi}{9} G \rho_{0, \BH} r_{c, \BH}^2 = \sigma_{c, \BH}^2}
\eq{\label{multimass:NcBH} N_{c, \BH}=0.98 \frac{\alpha_{1,\BH}}{\alpha_{2,\BH}} \lr{\frac{N_\BH}{10^2}}^{1/3}}

As in Section~\ref{ssec:sssc}, the number of binaries can be computed from the integral \eqref{singlemass:Nborig} of their binding energy distribution. To compute the number $N_{\BBH}$ of BBHs, we will use the form in \eqrefnp{singlemass:distribution} but changing the variables to their counterparts in the BH sub-cluster, e.g. $m\mapsto m_\BH$. This recovers the same form of the two-mass binary binding energy distribution from \cite{Retterer1980}, i.e.
\eq{N_{\BBH}\simeq 3.0\frac{G^3m_\BH^3n_{c, \BH}^2}{\sigma_{c, \BH}^8} V_{c, \BH} \abs{\phi_0}}
which we can express as 
\eq{\label{multimass:NbBH} N_{\BBH} = 2.8\frac{\alpha_{2,\BH}}{\alpha_{1,\BH}} \frac{\abs{\phi_0}}{\sigma_{c, \BH}^2} \lr{\frac{N_\BH}{10^2}}^{-1/3}}

In parallel to the result for the single-mass case, we obtain the prediction that the number of BBHs scales as $N_{\BBH}\propto N_\BH^{-1/3}$, although in this case the contribution of the $\abs{\phi_0}/\sigma_{c, \BH}^2$ term is larger. In the following section we will test the above equations against \nbody models.

\section{Description of the \NBODY models}
\label{section:nbody}
We simulate a set of single-mass and two-mass clusters and compare their global properties and the number of three-body binaries to the theory presented in the previous section. To isolate the dynamics of binary formation and disruption, we neglect  primordial binaries, stellar and binary star evolution, post-Newtonian physics, and the Galactic tidal field. We sample initial positions and velocities from a \citet{Plummer1911} model using the \textsc{McLuster} implementation from \citet{Kupper2011}. The exact choice of initial positions and velocities is not important for this study, as we will only consider the evolution beyond core collapse, after which the information about the initial conditions is erased \citep{LyndenBellEggleton1980}. Our simulations are run in H\'enon units,  i.e. $G=M_0=-4E_0=1$ \citep{1971Ap&SS..14..151H}, where $M_0$ is the initial cluster mass and $E_0$ is the initial energy of the cluster. For our assumption of virial equilibrium, the initial virial radius is $r_{v,0}=-GM_0^2/(2W_0)=1$, where $W_0=-1/2$ is the gravitational energy. The corresponding unit of time is $\tau_{dyn}=(r_{v,0}^3/GM_0)^{1/2}$.

We use the direct \nbody code \textsc{PeTar} \citep{Wang2020}, which is a high-performance code used for the modelling of large-$N$ collisional systems. It is based on a hybrid method: a Barnes-Hut \citep{BarnesHut1986} tree method for long-range forces and a few-body integrator for short-range interactions. The integrator for close interactions, \textsc{sdar} \citep{SDAR}, combines an explicit Hermite integrator for weakly perturbed binaries and a slow-down method for more compact subgroups.

The centre of the cluster as well as $r_c$ are defined following \cite{CasertanoHut1985, 1990ApJ...362..522M}, with $r_h$ determined from the bound stars in a reference frame where the density centre of the cluster is in the origin. We consider binaries as pairs of close stars whose binding energy is greater than $m\sigma_c^2$ -- which is proportional to the average kinetic energy of single stars in the core -- and that are bound to the cluster. 

\subsection{Single-mass models}
\label{section:nbody:singlemass}

We have run a family of single-mass models with logarithmically-spaced $N$, in the range $128 \leq N \leq 8192$, where each model is run several times for statistical significance. The input $N$ and number of runs are shown in Table~\ref{tab:nbodyparamssingle}. Within the aforementioned simplifications, single-mass models are scale-free, and thus $N$ is the only parameter that defines the post-collapse evolution of the cluster. The simulations are run through core collapse, which happens at $\tau_{cc}\simeq16 \tau_{rh, 0}$ for Plummer models \citep{Cohn1980}, with $\tau_{rh, 0}$ being the initial relaxation timescale \citep{SpitzerHart1971},
\eq{\label{singlemass:trh} \tau_{rh}=0.138 \frac{N}{\psi \ln \Lambda} \sqrt{\frac{r_h^3}{GM}},}
evaluated at the start of the simulation, where $\psi=1$ and $\ln\Lambda$ is the Coulomb logarithm with $\Lambda\simeq 0.11N$ for single-mass models \citep{GierszHeggie1994a}. The data for the post-collapse evolution is then taken starting from $t_0=1.1\tau_{cc}$ up until $t_f = t_0 + 20 \tau_{rh, 0}$, at constant intervals of 4 \nbody times. Values found for the cluster properties are averaged over all snapshots between the times $t_0$ and $t_f$, with their corresponding error bars being the standard deviation among different runs. The fittings are done with linear regressions, where we take the uncertainties into account by weighting the data points with the inverse squared value of their errors.

\begin{table}
\centering
\caption{Values of the input parameters in the \nbody runs of the single-mass clusters. Each row in the table represents a different cluster model, which is run multiple times (as shown in the last column) to obtain sufficient statistical significance.}
\begin{tabular}{@{}lrr@{}}
\toprule
$N$ & $t_f/\tau_{dyn}$ & Num. of runs\\ \midrule
128 & 143 & 64 \\
256 & 227 & 32 \\
512 & 376 & 16 \\
1024 & 642 & 8 \\
2048 & 1120 & 4 \\
4096 & 1986 & 4 \\
8192 & 3567 & 4 \\ \bottomrule
\end{tabular}
\label{tab:nbodyparamssingle}
\end{table}

\subsection{Two-mass models}
\label{ssec:nbody2}
For the two-mass \nbody models, we take our data from $t_0=1.1\tau_{cc}$ to $t_f=t_0 + 2 \tau_{rh, 0}$. We use a lower number of $\tau_{rh, 0}$ than in the single-mass models because in the two-mass models the binary phenomena happen within the BH subcluster and thus the relevant timescale is shorter. The impact of the adopted simulation time is discussed in Section~\ref{ssec:validity}. For the core-collapse timescale of two-mass clusters, we use the result of \cite{SungsooHyung1997}
\eq{\tau_{cc}=5.3 \lr{\frac{m_\star}{m_\BH}}^2\lr{\frac{N_\star}{N_\BH}}^{1/2}\tau_{rh,0},}
as well as their expression for $\tau_{rh, 0}$ (\eqrefnp{singlemass:trh}, but setting $\Lambda = 0.4 N$). The values of the three parameters $N_\BH$, $q$ and $\mu$ for the set of simulations that we have run is summarised in Table~\ref{tab:nbodyparams}. We set $N\in [5\times 10^4, 1\times 10^5, 2\times 10^5]$ and store the snapshots every 8 \nbody times. For the mass ratio $q$ we adopt $q\in[1/50, 1/25]$, to approximate metal-poor ($\log_{10}(Z/Z_\Sun)\simeq-1.5$) and metal-rich ($\log_{10}(Z/Z_\Sun)\simeq-0.5$) clusters, respectively. For the mass fraction $\mu$, we have chosen $\mu\in[0.025, 0.05, 0.1]$. Note that, for all combinations of these parameters, our clusters avoid gravothermal oscillations \citep{BreenHeggie2012a} and  satisfy the above criteria for Spitzer instability \citep{1969ApJ...158L.139S}, such that the evolution should be well-described by the theory of \citet{BreenHeggie2013}. 

\begin{table}
\centering
\caption{Values of the input parameters in the \nbody runs of the two-mass clusters. Each row in the table represents a different cluster model, which is run multiple times (as shown in the last column) to obtain sufficient statistical significance.}
\begin{tabular}{@{}llllr@{}}
\toprule
$N$ & $q$ & $\mu$ & $N_\BH$ & Num. of runs \\ \midrule
$5\times 10^{4}$ & 1/50 & 0.025 & 25 & 4 \\ 
$5\times 10^{4}$ & 1/25 & 0.025 & 50 & 4 \\ 
$5\times 10^{4}$ & 1/50 & 0.05 & 50 & 4 \\ 
$5\times 10^{4}$ & 1/25 & 0.05 & 100 & 4 \\ 
$5\times 10^{4}$ & 1/50 & 0.1 & 100 & 4 \\ 
$5\times 10^{4}$ & 1/25 & 0.1 & 200 & 4 \\ 
$1\times 10^{5}$ & 1/50 & 0.025 & 50 & 2 \\ 
$1\times 10^{5}$ & 1/25 & 0.025 & 100 & 2 \\ 
$1\times 10^{5}$ & 1/50 & 0.05 & 100 & 2 \\
$1\times 10^{5}$ & 1/25 & 0.05 & 200 & 2 \\ 
$1\times 10^{5}$ & 1/50 & 0.1 & 200 & 2 \\ 
$1\times 10^{5}$ & 1/25 & 0.1 & 400 & 2 \\ 
$2\times 10^{5}$ & 1/50 & 0.025 & 100 & 1 \\ 
$2\times 10^{5}$ & 1/25 & 0.025 & 200 & 1 \\ 
$2\times 10^{5}$ & 1/50 & 0.05 & 200 & 1 \\ 
$2\times 10^{5}$ & 1/25 & 0.05 & 400 & 1 \\ 
$2\times 10^{5}$ & 1/50 & 0.1 & 400 & 1 \\ 
$2\times 10^{5}$ & 1/25 & 0.1 & 800 & 1 \\   \bottomrule
\end{tabular}
\label{tab:nbodyparams}
\end{table}


\section{Comparing the steady post-collapse model to \NBODY simulations}

\label{section:comparing}

\subsection{Single-mass clusters}
\label{ssec:singlenb}
\begin{figure}
    \centering
    \includegraphics[width=\columnwidth]{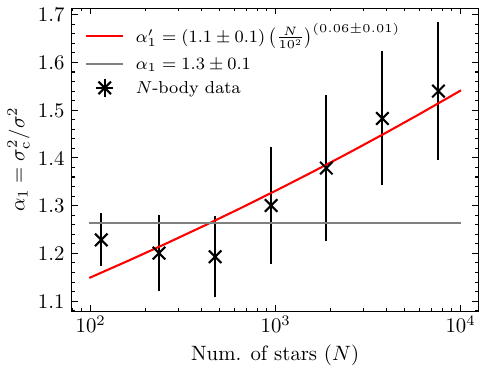}
        \caption{Ratio $\sigma_c^2/\sigma^2$ (\eqrefnp{singlemass:sigma2} and \eqrefnp{singlemass:sigma2c}) as a function of $N$ for the single-mass \nbody models (crosses with error bars). Fits for a constant $\alpha_1$ and an $N$-dependent $\alpha'_1$ are shown with grey and red lines, respectively.} 
    \label{fig:sigma2c}
\end{figure}

\begin{figure}
    \centering
    \includegraphics[width=\columnwidth]{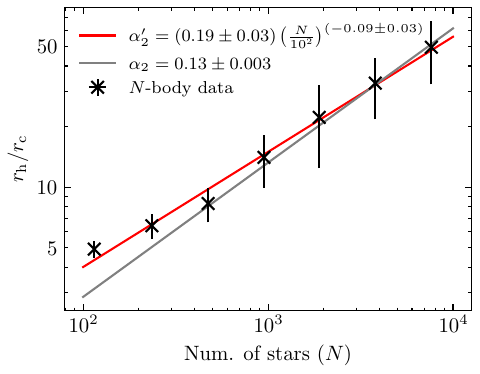}
    \caption{Ratio $r_h/r_c$ \eqref{singlemass:rhrc} as a function of $N$ for the single-mass \nbody models (crosses with error bars). Fits for a constant $\alpha_2$ and an $N$-dependent $\alpha'_2$ are shown with grey and red lines, respectively.} 
    \label{fig:rhrc}
\end{figure}

\begin{figure}
    \centering
    \includegraphics[width=\columnwidth]{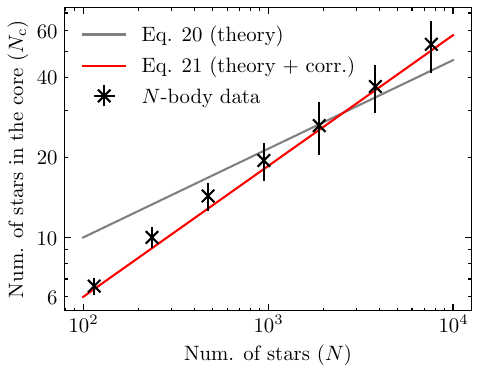}
    \caption{Number of stars in the cluster core as a function of $N$ for the single-mass \nbody models (crosses with error bars). Using $\alpha_{\it i}$ independent of $N$ (gray) we are able to roughly match the \nbody models, whereas using the $N$-dependant $\alpha'_{\it i}$ coefficients (red) yields a very accurate match.}
    \label{fig:Nc}
\end{figure}

\begin{figure}
    \centering
    \includegraphics[width=\columnwidth]{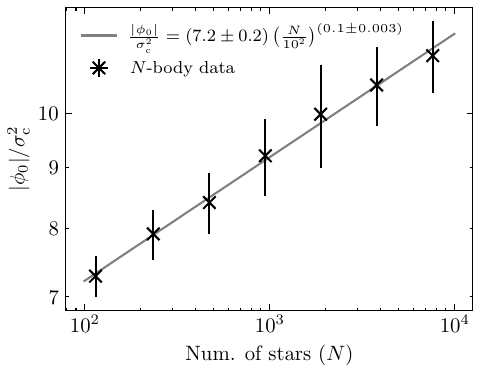}
    \caption{Ratio $\abs{\phi_0}/\sigma_c^2$ \eqref{singlemass:phicsigma2c} as a function of $N$ for the single-mass \nbody models (crosses with error bars).}
    \label{fig:phicsigma2c}
\end{figure}

\begin{figure}
    \centering
    \includegraphics[width=\columnwidth]{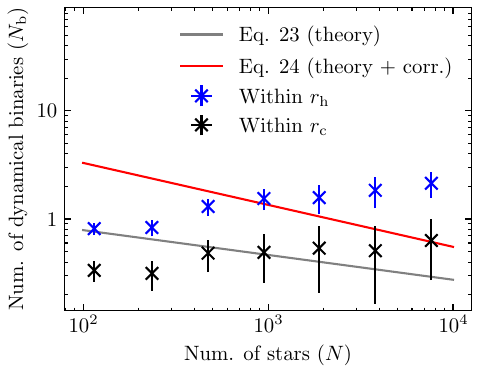}
    \caption{Number of binaries as a function of $N$ for the single-mass \nbody models (crosses with error bars). The theory over-predicts the number of binaries.}
    \label{fig:Nb}
\end{figure}

In this section we calibrate and compare the theoretical predictions of the single-mass model of Section~\ref{ssec:sssc} to the \nbody models of Section~\ref{section:nbody:singlemass}. We also allow the theory to include slight additional $N$-dependences in the $\alpha_1$, $\alpha_2$ parameters (that we label as $\alpha'_1$ and $\alpha'_2$, respectively). In Fig.~\ref{fig:sigma2c}, we show the ratio $\alpha_1=\sigma_c^2/\sigma^2$ as a function of $N$. We find that a constant $\alpha_1$ \eqref{singlemass:sigma2c} is a good approximation only for $N\lesssim10^3$, and $\alpha'_1$ increases slightly at larger $N$. A scaling $\alpha'_1 = 1.1 (N/10^2)^{0.06}$ describes the \nbody data well. This increase is likely due to the increase of $r_h/r_c$ with $N$, leading to a higher central density and dispersion, relative to the global values, for clusters with larger $N$. 

The $r_h/r_c$ ratio is shown Fig.~\ref{fig:rhrc}. The scaling of \eqrefnp{singlemass:rhrc} is a good approximation for constant $\alpha_2$ for $N\simeq 10^3 - 10^4$, whereas it deviates slightly at lower values of $N$, which, as discussed in Section \ref{ssec:sssc}, may be due to the weak $N$-dependence of the Coulomb logarithm in the denominator of the central relaxation timescale and in the $\abs{\phi_0}/\sigma_c^2$ term. A scaling of $\alpha'_2 = 0.19 (N/10^2)^{-0.09}$ provides a good fit to the \nbody data.

From these scaling laws, we are able to accurately predict the number of stars in the core (Fig.~\ref{fig:Nc}). The prediction for $N_c$ with constant $\alpha_{\it i}$ \eqref{singlemass:Nc} has the same form than the one given in \cite{Goodman1984},
\eq{\label{singlemass:Nc_param} N_c=10 \lr{\frac{\alpha_1}{1.3}} \lr{\frac{\alpha_2}{0.13}}^{-1} \lr{\frac{N}{10^2}}^{1/3},}
which yields a rough estimate of $N_c$ from the \nbody models within <50\% over two orders of magnitude in $N$. Using the $N$-dependent expressions for $\alpha'_1$ and $\alpha'_2$ yields
\eq{\label{singlemass:Nc_corr} N_c'=0.58\lr{\frac{N}{10^2}}^{0.17} N_c}
which accurately describes the \nbody data. 
Furthermore, in Fig.~\ref{fig:phicsigma2c} we show the ratio $|\phi_0|/\sigma_c^2$ as a function of $N$, which can be approximated by the power-law relation
\eq{\label{singlemass:phicsigma2c} \frac{\abs{\phi_0}}{\sigma_c^2}\simeq 7.2 \left(\frac{N}{10^2}\right)^{0.1}.}

Now all ingredients of the expression for the number of binaries have been discussed, it is time to combine them and compare to $N_b$ we find in the simulations. We obtain
\eq{\label{singlemass:Nb_param} N_b = 1.9\left(\frac{\alpha_2}{0.13}\right)\left(\frac{\alpha_1}{1.3}\right)^{-1}\left( \frac{\abs{\phi_0}/\sigma_c^2}{7.2}\right) \left(\frac{N}{10^2}\right)^{-1/3}}
Which can be corrected by the deviations from the scaling laws above to yield
\eq{\label{singlemass:Nb_corr} N'_b = 1.7 \left(\frac{N}{10^2}\right)^{-0.17} N_b, }
such that a cluster with $\sim100$ stars is expected to have $\sim3$ binaries.

In Fig.~\ref{fig:Nb} we show both the prediction for $N_b$ and the \nbody results. The prediction \eqref{singlemass:Nb_corr} approximately reproduces the number of binaries in the core at large $N$ ($N\gtrsim10^3$), whereas in the smallest clusters $N_b$ is over-predicted by an order of magnitude. It could be reasonable to consider binaries outside of the core, as the interactions with single stars increase the apocentre of their orbits as they evolve in their lifecycle. When we consider all binaries within $r_h$, the number of binaries in the \nbody models is higher, and increases slightly with $N$, that is, the opposite $N$-dependence. The simulations are compatible with the clusters only having a single central binary at any point in time, which spends a fraction of its lifetime outside the core.

\subsection{Two-mass clusters}
\label{ssec:twonb}

\begin{figure}
    \centering
    \includegraphics[width=\columnwidth]{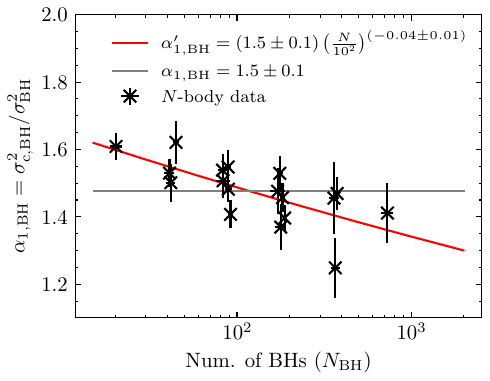}
    \caption{Ratio $\sigma_{c, \BH}^2/\sigma_\BH^2$ \eqref{multimass:sigma2cBH} as a function of $N_\BH$ for the two-mass \nbody models (crosses with error bars). Fits for a constant $\alpha_{1, \BH}$ and an $N$-dependent $\alpha'_{1, \BH}$ are shown with grey and red lines, respectively. }
    \label{fig:sigma2cBH}
\end{figure}

\begin{figure}
    \centering
    \includegraphics[width=\columnwidth]{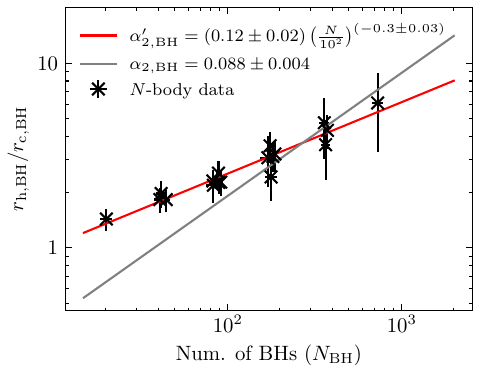}
    \caption{Ratio $r_{h,\BH}/r_{c, \BH}$ \eqref{multimass:rhBHrcBH} as a function of $N_\BH$ for the two-mass \nbody models (crosses with error bars). Fits for a constant $\alpha_{2, \BH}$ and an $N$-dependent $\alpha'_{2, \BH}$ are shown with grey and red lines, respectively. }
    \label{fig:rhBHrcBH}
\end{figure}

In this section we compare the theoretical prediction of the two-mass model of Section~\ref{ssec:sstc} to the \nbody models of Section~\ref{ssec:nbody2}. The velocity dispersion within the BH sub-cluster's core is shown in Fig.~\ref{fig:sigma2cBH}, where we find reasonable agreement for a constant $\alpha_{1, \BH}$ \eqref{multimass:sigma2cBH}, but a slight decline of $\alpha_{1, BH}$ with $N_\BH$ is preferred, equivalent to the single-mass case. For the ratio $r_{h,\BH}/r_{c, \BH}$ in the BH sub-cluster (Fig.~\ref{fig:rhBHrcBH}), we find that the scaling law of \eqrefnp{multimass:rhBHrcBH} is a valid approximation to the data. Similarly to the single-mass case, the deviation at low $N_\BH$ can be attributed to the weak $N_\BH$ dependency of the Coulomb logarithm. However, in the two-mass case the deviation is larger due to the smaller number of BHs in the core.

\begin{figure}
    \centering
    \includegraphics[width=\columnwidth]{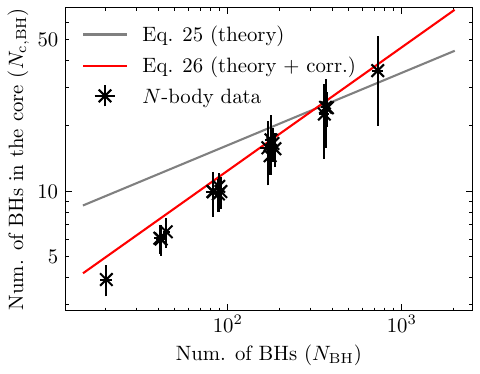}
    \caption{Number of BHs in the sub-cluster core as a function of $N_\BH$ for the two-mass \nbody models (crosses with error bars). Using $\alpha_{{\it i}, \BH}$ independent of $N_\BH$ (gray) we are able to roughly match the \nbody models, whereas using the $N_\BH$-dependant $\alpha'_{{\it i}, \BH}$ coefficients (red) yields a good match.}
    \label{fig:NcBH}
\end{figure}

From these relations, we can predict the number of BHs in the sub-cluster core (Fig.~\ref{fig:NcBH}). As in the single-mass case, constant $\alpha_{{\it i}, \BH}$ \eqref{multimass:NcBH} give a rough approximation to $N_{c, \BH}$,
\eq{\label{multimass:Nc_param} N_{c, \BH} = 16 \lr{\frac{\alpha_{1, \BH}}{1.5}} \lr{\frac{\alpha_{2, \BH}}{0.088}}^{-1} \lr{\frac{N_\BH}{10^2}}^{1/3}}
with the discrepancy being largest at lower $N_\BH$. Using the corrected $\alpha'_{{\it i}, \BH}$ yields 
\eq{\label{multimass:Nc_corr} N'_{c, \BH} = 0.76\lr{\frac{N_\BH}{10^2}}^{0.23} N_{c, \BH},}
which accurately reproduces the results from the \nbody models for $N_\BH \gtrsim 50$, although it slightly overpredicts the results for the sub-clusters with the fewest black holes in the core ($N_{c, \BH} \lesssim 7$).

\begin{figure}
    \centering
    \includegraphics[width=\columnwidth]{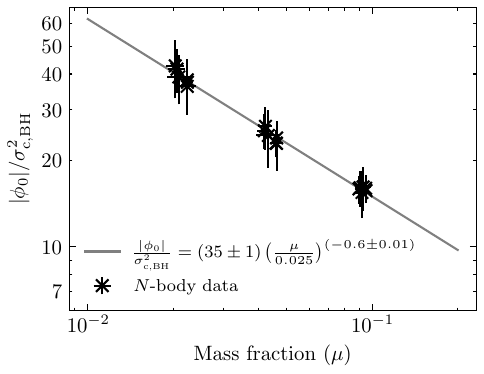}
    \caption{Ratio $\abs{\phi_0}/\sigma_{c, \BH}^2$ \eqref{multimass:phicsigma2cBH} as a function of $\mu$ for the two-mass \nbody models (crosses with error bars).}
    \label{fig:phicsigma2cBH}
\end{figure}

The above scaling laws are complemented by a fit to $\abs{\phi_0}/\sigma_{c, \BH}^2$ (Fig.~\ref{fig:phicsigma2cBH}). From the three parameters of the model, we find that this value only correlates strongly with $\mu$. The best-fit power law is
\eq{\label{multimass:phicsigma2cBH} \frac{\abs{\phi_0}}{\sigma_{c, \BH}^2}\simeq 35 \left(\frac{\mu}{0.025}\right)^{-0.62}.}
Although this equation may be taken to imply a very deep central potential, this is not the case. The apparently large value is an artefact of expressing the potential in units of the central velocity dispersion of the BHs. Using \citet[equation 4]{BreenHeggie2013}, we can write it in terms of the central (total) velocity dispersion, yielding a much smaller value
\eq{\frac{\abs{\phi_0}}{\sigma_c^2}\simeq 4 \lr{\frac{\mu}{0.025}}^{-0.22} \lr{\frac{q}{1/25}}^{0.4}.}

Using the calibrations for the scaling laws (equations~\ref{multimass:sigma2cBH}-\ref{multimass:NcBH}), we can evaluate the prediction for the number of BBHs of \eqrefnp{multimass:NbBH} to obtain
\eq{\label{multimass:Nb_param} N_{\BBH} = 6 \left(\frac{\alpha_{2, \BH}}{0.088}\right)\left(\frac{\alpha_{1, \BH}}{1.5}\right)^{-1}\left( \frac{\abs{\phi_0}/\sigma_{c, \BH}^2}{35}\right)\left(\frac{\mu}{0.025}\right)^{0.62} \left(\frac{N_\BH}{10^2}\right)^{-1/3}.}
This can be corrected with the deviations from the scaling laws as
\eq{\label{multimass:Nb_corr} N'_{\BBH}=1.3 \left(\frac{N_\BH}{10^2}\right)^{-0.23} N_{\BBH}.}

\begin{figure}
    \centering
    \includegraphics[width=\columnwidth]{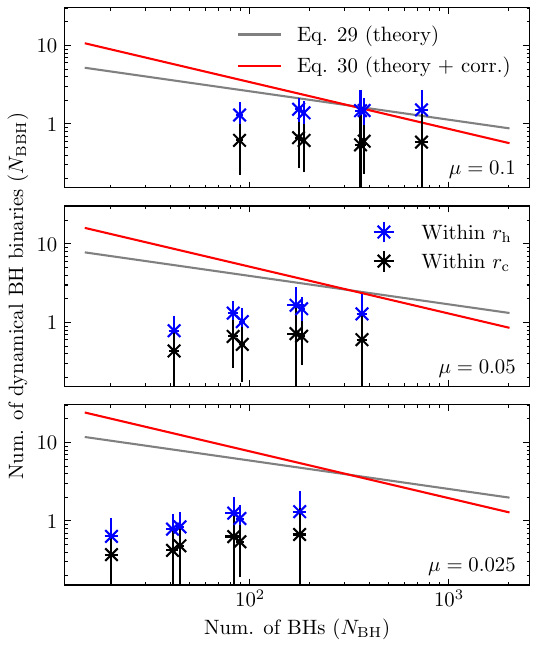}
    \caption{Number of BBHs as a function of $N_\BH$ and $\mu$ for the two-mass \nbody models (crosses with error bars). The number of BBHs is roughly constant and the theory vastly overpredicts the number of binaries.}
    \label{fig:NbBH}
\end{figure}

Although the two-mass steady post-collapse model is able to describe the other properties of the \nbody model, it over-predicts the number of BBHs (Fig.~\ref{fig:NbBH}). This result is very similar to the single-mass case, although in this case the discrepancy is larger. In general, the discrepancy is most notable in models with smaller $\mu$ and smaller $N_\BH$.

\section{The cause of the dearth of binaries: binary-binary interactions}
\label{section:binbininteractions}
\label{section:binBHbinBHinteractions}
In the previous sections we find that the theoretical predictions for the population of binaries are unable to describe the number of three-body binaries in the \nbody runs, even if all the other elements of the model do match the simulations. This discrepancy had already been observed in \cite{GierszHeggie1994b}, but no cause was definitively identified. In this section we search for an explanation for the discrepancy between the theory and the \nbody models. Because binary formation is well understood \citep{Heggie1975}, we should search for a missing ingredient in the model for binary evolution. The models of \citet{Retterer1980} and \citet{GoodmanHut1993} assume that the binding energies of binaries evolve because of interactions between binaries and single stars, and binary-binary interactions are ignored with the argument that they are rare. However, they start to play a role if the binary fraction in the core is above some critical value. The most likely outcome of a binary-binary interaction is that at least one of the binaries is ionised ($\sim 95\%$ of the outcomes for the scattering experiments in \citealt{AntogniniThompson2016}), which can explain the dearth of binaries. Because the total number of binaries in the \nbody models is $\mathcal{O}(1)$, ignoring interactions among binaries is a justified assumption for clusters with $N_c\gg 1$. However, as we have shown in Fig.~\ref{fig:Nc}, $N_c\lesssim 20$ for $N\lesssim 10^3$, whilst the theory predicts $N_b\gtrsim 2$. In fact, the predicted binary fraction in the core is $f_b= N_b/N_c\simeq 0.6 (N/10^2)^{-0.9}$. Therefore, if the model for the binary energy distribution in the steady post-collapse model is correct, then binary-binary interactions would be important, so we need to understand their effect on shaping the energy distribution of binaries. 

We start by estimating the critical binary fraction in the core, above which binary-binary interactions become more frequent than binary-single interactions. The interaction cross-section for a binary with a single goes as $\Sigma_{bs}\propto G r_p m_{tot}/v_{bs}^2$, where $v_{bs}$ is the relative velocity, $r_p\simeq a$ is the relevant minimum distance (with $a$ the semimajor axis of the binary), and $m_{tot}=3m$ is the total mass of the interacting binary and single.
Assuming equipartition, the typical relative velocity among binaries, $v_{bb}$, is a factor of $\sqrt{3/4}$ lower than among binaries and singles ($v_{bb}=\sqrt{3/4}v_{bs}$). Then, taking $m_{tot}=4m$ and $r_p=2a$ for binary-binary interactions, we estimate that the cross section for these is $\Sigma_{bb} \simeq3.6\Sigma_{bs}$. The ratio of the rate of binary-binary interactions to the rate of binary-single interactions, $\Gamma_{bb}/\Gamma_{bs}$, is
\eq{\frac{\Gamma_{bb}}{\Gamma_{bs}}\simeq \frac{N_b^2\Sigma_{bb}v_{bb}}{N_bN_s\Sigma_{bs}v_{bs}}\simeq 3.1 f_b }
This estimate shows that binary-binary interactions are expected to be more frequent than binary-single interactions for\footnote{This derivation assumes a cluster with no primordial binaries or triples. See \citet{LeighSills2011} for a discussion of the general case.} $f_b\gtrsim 0.3$ in the core. Although this may seem high, the core contains of order 10 stars, so this criterion is already met if 3 binaries are created. This is approximately the number of hard binaries predicted by \citet{GoodmanHut1993} for clusters $N\simeq100$, and we note that their models predict an even higher formation rate of soft binaries which have a larger cross section for interactions and can therefore also contribute to binary ionisation in a way that is not captured by the \citet{GoodmanHut1993} model. We will therefore test whether binary-binary interactions can reduce the number of binaries seen in the \nbody models.

\subsection{Single-mass models}
\label{ssection:binbininteractions}

\begin{figure}
    \centering
    \includegraphics[width=\columnwidth]{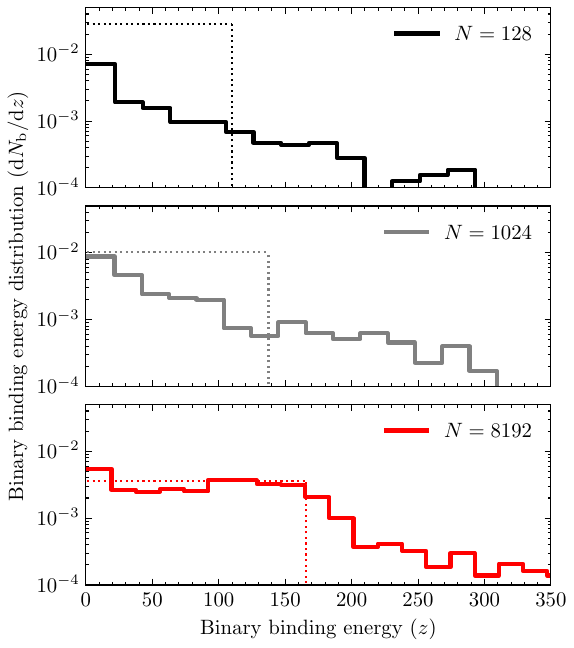}
    \caption{Distribution of the binding energies of the binaries as a function of $N$ for the single-mass \nbody models. The theory (dotted lines) does not match the observed distribution (full lines), with the discrepancy being highest at lower $N$ and higher $z$. }
    \label{fig:z_step}
\end{figure}

As stated above, the most likely outcome of a binary-binary interaction is that at least one of the binaries is ionised. If the timescale for this ionisation mechanism is shorter than the time needed to form new binaries, then the predicted uniform binding energy distribution \eqref{singlemass:distribution} is never realised.  Instead, one would find a decline in binaries at higher $z$, because they are destroyed at some point during their life-cycle. Furthermore, this deviation should be larger at smaller $N$, where the theory predicts the highest binary fraction. In Fig.~\ref{fig:z_step} we show the histogram of binary binding energies within the core during the steady post-collapse evolution  of three cluster models. As predicted, we observe that the predicted uniform distribution is not realised, supporting the idea that binaries are destroyed before they reach the maximum binding energy.

\begin{figure}
    \centering
    \includegraphics[width=\columnwidth]{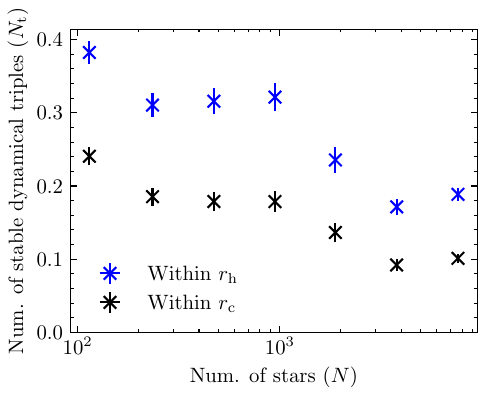}
    \caption{Number of dynamically-formed stable triples $N_t$ as a function of $N$ (crosses with error bars) in the single-mass \nbody model. The non-zero value of $N_t$ implies the presence of binary-binary interactions.}
    \label{fig:Nt}
\end{figure}

The qualitative difference of binary-binary interactions with respect to binary-single interactions is the possibility of formation of stable triple systems \citep{Zevin2019}. Although the formation of stable triples via binary-single interactions is not energetically forbidden, the probability of such process is zero \citep[p. 211]{HeggieHut2003}. Thus, the existence of dynamically-formed stable triple systems is a necessary (but not sufficient,  Section~\ref{sssec:altpath}) condition to confirm the importance of binary-binary interactions in a cluster. The formation of triples can easily be measured in an \nbody model, where we identify triples as bound states of three stars that verify the stability criterion from \cite{MardlingAarseth2001}. The results for the average number of stable, dynamically-formed triples $N_t$ in the single-mass model is shown in Fig. \ref{fig:Nt}, where we see that not only such triples are present, but their number is an inverse function of the number of stars of the cluster.

In order to gauge the impact of binary-binary interactions in the binary population, we will estimate the relative importance of the triple formation rate --  assuming the theoretically expected binary distribution function in \eqrefnp{singlemass:distribution} -- with respect to the net binary formation rate $\Gamma_b$ \eqref{singlemass:cbar}. For triple formation in binary-binary interactions to be negligible, this ratio should be much smaller than one. We can estimate the triple formation rate with the corresponding cross sections. We will use the results of \cite{AntogniniThompson2016}, where the authors considered scattering experiments of equal-mass, circular binaries in the Newtonian point-particle limit. Their quoted result for the cross section of stable triple formation is
\eq{\Sigma(a_1, a_2, v) = \pi \lr{a_1^2+a_2^2}\lr{\frac{\hat{v}^2}{0.567}+\frac{\hat{v}^6}{0.0279}}^{-1}}
The cross section depends on the semimajor axes $a_1$, $a_2$ of the incoming binaries and the ratio $\hat{v}$ of their relative velocity to their critical velocity, $\hat{v}=v/\sqrt{Gm(a_1+a_2)/(a_1a_2)}$. The total rate of triple formation in binary-binary encounters per unit volume can be obtained in the `n-sigma-v' formalism by the integration of these variables using, respectively, the distribution of $z$ \eqref{singlemass:distribution} and the distribution for the relative velocities. If we assume that the distribution of velocities in the core is Maxwellian with a dispersion of $\sigma_c$, then the dispersion of relative velocities among singles is $\sqrt{2}\sigma_c$. If we then take binaries to be twice as massive and assume equipartition between singles and binaries, we obtain that the relative velocities of binaries also follow a Maxwellian distribution with a dispersion of $\sigma_c$ ($f_{Maxwell}(v)\propto v^2/\sigma_c^3 \exp(- 0.5v^2/\sigma_c^2$)). Therefore, the triple formation rate is
\eq{\label{binbin:gamma}\Gamma_{t, bb} = \iiint \frac{f(z_1)f(z_2)}{2}f_{Maxwell}(v) \Sigma(z_1, z_2, v) v \diff z_1\diff z_2 \diff v}

\begin{figure}
    \centering
    \includegraphics[width=\columnwidth]{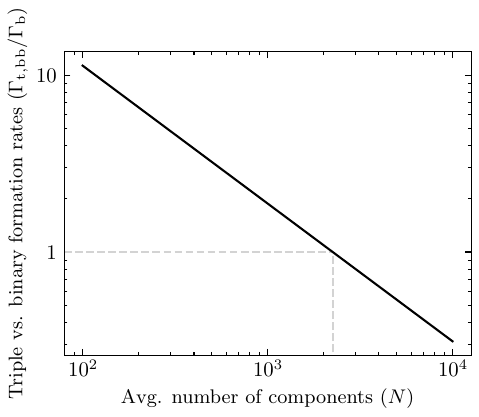}
    \caption{Ratio of the triple formation rate $\Gamma_{t,bb}$ to the binary formation rate $\Gamma_b$. This ratio is $\gg 1$ for $N\lesssim 10^3$, which confirms the importance of binary-binary interactions in that $N$ regime. The ratio scales as $\Gamma_{t,bb}/\Gamma_b\propto N^{-0.8}$.}
    \label{fig:rate}
\end{figure}

The ratio $\Gamma_{t, bb}/\Gamma_b$ is then expressed only as a function of $N$ by using the scaling relations in Table~\ref{tab:scalinglaw} and shown in Fig.~\ref{fig:rate}. The ratio is greater than one for $N\lesssim 2000$ and declining approximately as $N^{-0.8}$. This supports the idea that binary-binary interactions are destroying three-body binaries faster than they form, and that this effect is especially relevant in lower-mass clusters. In turn, this gives a theoretical explanation for the observed presence of a single three-body binary in \nbody simulations, as every time a new binary forms it is rapidly destroyed in a binary-binary interaction.

\subsection{Two-mass models}
In section \ref{ssection:binbininteractions} we showed that, for single-mass models, binary-binary interactions are non-negligible even when no primordial binaries are present. We will now argue that this is a general prediction that also applies to the population of BBHs in two-mass models.

\begin{figure}
    \centering
    \includegraphics[width=\columnwidth]{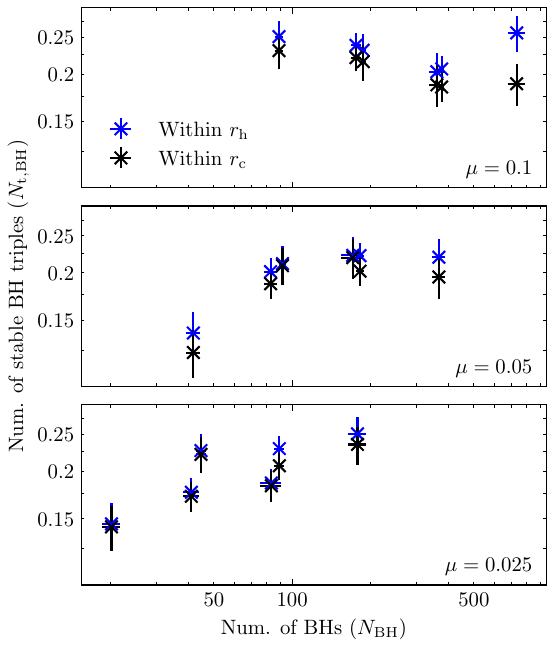}
    \caption{Number of dynamically-formed stable BH triples $N_{t, \BH}$ as a function of $N_\BH$ and $\mu$ (crosses with error bars) in the two-mass \nbody model. The non-null value of $N_{t, \BH}$ implies the presence of BBH-BBH interactions.}
    \label{fig:NtBH}
\end{figure}

\begin{figure}
    \centering
    \includegraphics[width=\columnwidth]{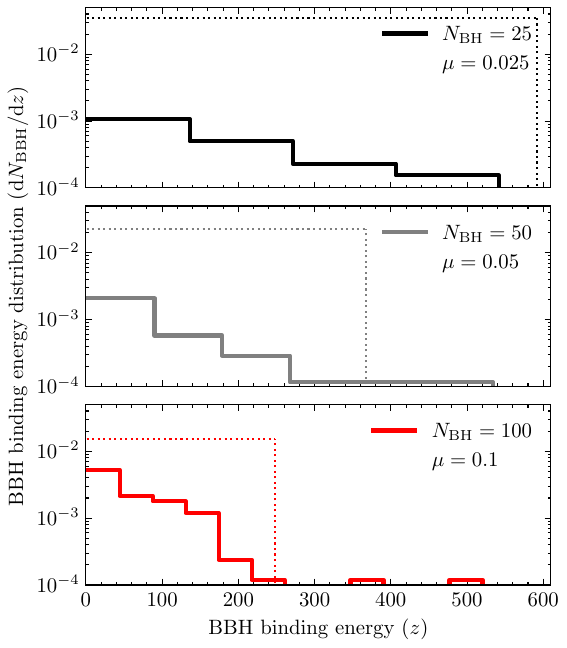}
    \caption{Distribution of the binding energies of the BBHs as a function of $N_\BH$ and $\mu$ for the two-mass \nbody models with $N=50000$ and $q=1/50$. The theory (dotted lines) does not match the observed distribution (full lines), with the discrepancy being highest at lower $N_\BH$, $\mu$ and higher $z$. }
    \label{fig:z_step_2m}
\end{figure}

As before, we show in Fig.~\ref{fig:z_step_2m} that a uniform BBH binding energy distribution is not realised, with the discrepancy being greatest at lower $N_\BH$ and higher $z$. Contrary to the single-mass case, here the models with lower $N_{BH}$ have higher $z_{ej}$ because of their smaller $\mu$ (see equation~\ref{multimass:Nb_param}).
Furthermore, we show that BBH-BBH interactions happen by looking at the formation of stable triple BH systems. Indeed, in Fig.~\ref{fig:NtBH} we can see that stable BH triple formation happens for all clusters, independently of their number size, BH mass fraction or mass ratio. In the two-mass case, the values of $N_\BH$ correspond to the smaller end of the range of $N$ in the single-mass case, so we do not observe the decrease in the number of triples that is seen at large $N$ in the single-mass models. The presence of stable triples in \nbody models without primordial binaries had already been noticed before, e.g. \cite{Aarseth2012, Banerjee2018b}. Although stable with respect to their internal kinematics, the triples that form in our models are rather soft and can easily be destroyed via interactions with unbound singles.

\begin{figure}
    \centering
    \includegraphics[width=\columnwidth]{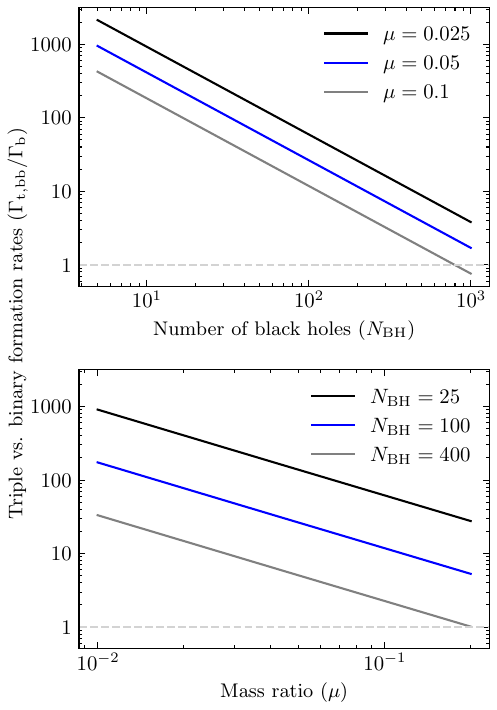}
    \caption{Dimensionless ratio of the BH triple formation rate $\Gamma_{t, bb}$ to the BBH formation rate $\Gamma_b$ as a function of the number of BHs $N_\BH$ (top panel) and the mass fraction $\mu$ (bottom panel). This ratio is much greater than one, which points towards the non-negligible relevance of BBH-BBH interactions when modelling the three-body BBH population. The ratio scales approximately as $\Gamma_{t, bb}/\Gamma_b\propto \mu^{-1.2} N_{\BH}^{-1.2}$.}
    \label{fig:multimass:binbinrate}
\end{figure}

In parallel to our argument in the previous section, we evaluate the rate of BBH-BBH interactions. We use the same form than in \eqrefnp{binbin:gamma}, although we will now use the scaling relations in equations~(\ref{multimass:sigma2cBH}-\ref{multimass:NcBH}) to obtain the ratio of rates as a function of $N_\BH$ and $\mu$. This ratio of rates is shown in Fig.~\ref{fig:multimass:binbinrate}. As can be seen, the ratio is much larger than one for all values of the number of BHs, mass ratio or mass fraction, which explains the discrepancy between the observed and predicted number of binaries in Fig.~\ref{fig:NbBH}. We can therefore conclude that BBH-BBH interactions are a key ingredient to shape the distribution of three-body BBHs. In the next section we discuss possible caveats in our arguments and alternative interpretations.

\section{Discussion and conclusions}
\label{sec:discussion}
\subsection{Validity of the assumptions}
\label{ssec:validity}
In order for the binaries to populate the binary binding energy distribution in the absence of binary-binary interactions (Fig.~\ref{fig:z_step} and Fig.~\ref{fig:z_step_2m}), the properties of the cluster must evolve slowly with time. The criterion for this is that the timescale at which the cluster expands, $\tau_{exp}$, is larger than the binary lifecycle, $\tau_{bin}\simeq (z_{ej}-z_h) /\dot{z}\simeq z_{ej}/\dot{z}$. We can define the timescale for expansion similarly to $\tau_{bin}$ by assuming that $\tau_{exp}\simeq r_h/\dot{r}_h$. Using H\'{e}non's principle \citep{Henon1975}, we can relate the energy production of the core to its global properties as 
\eq{\dot{E} \simeq \zeta \frac{|E|}{\tau_{rh}}}
with $\zeta\simeq 0.1$ \citep{1965AnAp...28...62H,AlexanderGieles2012}. If we assume that the cluster energy scales as $E\propto GM^2/r_h$ and that the mass loss is negligible, we obtain
\eq{\label{disc:tauexp} \tau_{exp}=\frac{\abs{E}}{\dot{E}_h}\simeq \frac{\tau_{rh}}{\zeta}. }
This can be compared to the binary lifecycle, which we find from the scaling laws of Section~\ref{ssec:singlenb} 
\eq{\label{discussion:taubin}\tau_{bin}&\simeq \frac{z_{ej}}{\dot{z}}\simeq 300 \lr{\frac{\psi \log \Lambda}{10}}\lr{\frac{N}{10^2}}^{-1.2}  \tau_{rh},\nonumber\\&\simeq 300  \lr{\frac{N}{10^2}}^{-0.2}  \tau_{dyn}.} 
We also give the value of this timescale in terms of the dynamical time $\tau_{dyn}$ by using that $\tau_{rh}/\tau_{dyn}\simeq 0.1 N/(\psi \log \Lambda)$ \citep{HeggieHut2003}. The binary timescale is approximately a constant number of \nbody times, weakly decreasing with $N$. This somewhat counter-intuitive result is a consequence of the larger central concentration of high-$N$ models, leading to a faster evolution of the binaries. Alternatively, if we had considered homologous evolution ($r_c\propto r_h$ and $N_c \propto N$), this timescale would increase strongly with $N$, $\tau_{bin} \propto N^2 \tau_{dyn}$.

For the single-mass case, we find $\tau_{exp}>\tau_{bin}$ for $N\gtrsim 800$, so we are justified in taking the slow evolution approximation in larger clusters. For smaller clusters, the expansion leads to a decrease in the binding energy distribution, as \eqref{singlemass:distribution}
\eq{f(z)\propto \frac{m^3n_c^2}{\sigma_c^6}\propto \frac{r_h^{3}N_c^2}{N^3r_c^{6}}\propto N^{5/3}r_h^{-3}}
This implies that, for a fixed $N$, the expansion of the cluster would lead to a lowering of the binding energy distribution with time, so we would find a pile-up of binaries at large $z$, because these were formed earlier when $f(z)$ was higher. We actually see a lack of binaries at large $z$, making the absence of binaries even more important for $N\lesssim800$.

Another timescale to consider is the duration of our simulations, $t_f-t_0$. If this timescale is much shorter than the binary lifetime $\tau_{bin}$, the binaries would not have enough time to populate the binding energy distribution, even in the absence of binary-binary interactions. The duration of the simulations is $t_f-t_0=20 \tau_{rh, 0}$. Because the cluster expands, the actual number of elapsed relaxation times is less than 20. We can estimate the instantaneous relaxation timescale by assuming that \citep{1965AnAp...28...62H} 
\eq{\tau_{rh}(t)=\begin{cases}
\tau_{rh, 0}&t<\tau_{cc}\\
(t/\tau_{cc})\tau_{rh, 0}&t>\tau_{cc}
\end{cases}}
So we obtain
\eq{t_f-t_0=\langle\tau_{rh}\rangle\int^{t_f}_{t_0}\frac{\diff t}{\tau_{rh}(t)}=12 \langle\tau_{rh}\rangle}
which we can compare to equation~(\ref{discussion:taubin}) to show that $t_f-t_0>\tau_{bin}$ for $N\gtrsim 380$, so this condition is fulfilled in all models but the least massive ones. A similar calculation for the two-mass models yields (ignoring the variation of the Coloumb logarithms)
\eq{\frac{t_f-t_0}{\tau_{bin}} \simeq 0.3 \lr{\frac{N_{BH}}{10^2}}^{0.8}\lr{\frac{q}{1/50}}^{1.1} \lr{\frac{\mu}{0.025}}^{-1.3}}
which is larger than unity for models with $N_{BH}\gtrsim 150\ (450)$ for $q=1/25\ (1/50)$ and $\mu=0.025$. This shows that our two-mass models may not have been evolved for long enough to be comfortably in the regime where the binding energy distribution is well populated and so may explain why we do not see an obvious decrease in the number of stable triples with $N_{BH}$ in Fig.~\ref{fig:NtBH} as in the single-mass case. To check whether there is evolution of the binary energy distribution, we compared $f(z)$ in the first half to $f(z)$ in the second half of the simulation interval. The two distributions are statistically indistinguishable for all $N_{BH}$, suggesting that the simulation time is long enough for $f(z)$ to have reached its steady post-collapse shape.

Yet another possible caveat is the convergence of the data points to the true underlying distribution. For low-$N$ models, we have many runs and therefore the convergence is attained, but for the most massive models, we have only a handful of runs (see Table~\ref{tab:nbodyparamssingle}). In this case, the binary lifetime is longer than the \nbody simulation timescale and thus convergence is attained by combining multiple snapshots from the few runs.

\subsection{Alternative pathways towards stable triple formation: `binary-single-single triple formation'}
\label{sssec:altpath}
Binary-binary interactions are not the only mechanism towards stable triple formation. Alternatively, one could consider triple formation in a binary-single-single interaction, in the same way as three-body binary formation, but with one component being a pre-existing binary. To estimate the rate of hard binary formation in three-body encounters, \cite{HeggieHut2003} consider the rate at which two unbound stars approach closer than $a_h$ and multiply that rate by the probability that a third star is present within the volume, which is roughly $n_ca_h^3$. We can estimate the rate of triple formation due to binary-single-single interactions, $\Gamma_{t, bss}$, by a similar argument, although performing the change $n_c\mapsto n_b$ and $a_h \mapsto 3 a_h$ (so that the outer orbit of the triple is about $\sim 3 a_h$). Then, the ratio of the rate of triple production over binary formation is roughly $\Gamma_{t, bss}/\Gamma_{b}\sim n_b (3a_h)^3/(n_ca_h^3)=27/N_c\simeq3(N/10^2)^{-1/3}$, where we use the scaling $N_c(N)$ from equation~(\ref{singlemass:Nc_param}) and, as seen in the \nbody models, we assume that there is a single binary in the core. Therefore, the production of triples due to binary-single-single interactions could be of similar importance as binary-binary interactions. This applies to soft triples with large outer semimajor axis of $3a_h$. The formation rate of `hard' triples is a factor of 27 lower. The $N$-dependence in $\Gamma_{t,bb}/\Gamma_b$ is steeper ($N^{-0.8}$, see Figs.~\ref{fig:rate} and \ref{fig:multimass:binbinrate}) so we conclude that binary-single-single triple formation may occur, but that triple formation via binary-binary encounters dominates for the smallest-$N$ clusters. In addition, we showed in Fig.~\ref{fig:multimass:binbinrate} that triple formation in binary-binary encounters is more important in two-mass models, whereas the above scaling for binary-single-single triple formation should be the same for two-mass models, but with $N$ replaced by $N_{BH}$.

\subsection{The effect of galactic tides}
So far we have neglected the effect of the galactic tidal field. Here, we discuss how our results would be affected by the inclusion of galactic tides. We start by considering the flow of energy through the cluster's half-mass radius, which is governed by the coefficient $\zeta$ \eqref{disc:tauexp}. It is shown in \citet{Gieles2011} that the value of $\zeta$ in tidally-limited clusters and isolated clusters is equal to within 20\%, and that  the first half of evolution is similar in tidally-limited and isolated clusters. Furthermore, the authors estimate that two thirds of Milky Way globular clusters are in this first expansion phase. Per H\'enon's principle, this energy flow is balanced by the production of energy in the core, and therefore the density profile within a few $r_h$ and the overall results of this work are independent of the underlying host galaxy potential. The only difference that the tidal stripping introduces is the possibility of evaporation pathways where the mass loss rate of stars is higher than the mass loss rate of BHs, such that the cluster evolves to a 100\% BH cluster \citep[$\mu\rightarrow1$, see][]{2011ApJ...741L..12B,Gieles2021}. As stated in Section~\ref{ssec:sstc}, we do not consider high-$\mu$ clusters, although we roughly expect them to behave like the single-mass models in the limit $\mu\rightarrow 1$. Nevertheless, observations point towards most clusters in the Milky Way having a small BH mass fraction of less than 1\% \citep{Dickson2023}.

\subsection{Implications for GW astronomy}

In this work, we did not include post-Newtonian terms; however, here we discuss how the high rate of binary-binary encounters (Section \ref{section:binbininteractions}) and the resulting stable triples could lead to GW inspirals by captures. These captures happen when two BHs have a close approach, radiate away orbital energy and merge in a short timescale due to the emission of GWs \citep{Samsing2018}. According to the scattering experiments by \cite{Zevin2019}, binary-binary interactions are roughly five times more likely than binary-single interactions to produce such mergers due to their more complex resonant intermediate states. 

The binary-binary interactions described in this paper have at least one binary near the hard-soft boundary. During a resonant intermediate state, two of the BHs can enter a highly eccentric orbit that ends in a GW capture. This orbit, therefore, starts with a large semimajor axis ($a\sim a_h$), which requires a nearly radial orbit ($e\sim1$) to trigger a GW capture. We can use the capture criterion by \citet{Samsing2018} to quantify what the eccentricity, $e$, must be
\eq{1-e\lesssim \lr{\frac{4.6 G m}{a c^2}}^{5/7}.}
To put it in context, for a $10^5 \text{ M}_\odot$ cluster with $r_h=1 \text{ pc}$, a capture between two $10 \text{ M}_\odot$ BHs would start at $\sim 60 \text{ AU}$ and $1-e \sim 10^{-6}$. As the binary inspirals, this eccentricity may be radiated away. To estimate this, we find the peak GW frequency $f$ from \citet{Wen2003},
\eq{f = \frac{1}{\pi}\sqrt{\frac{2Gm_\BH}{a^3}}\frac{(1+e)^{1.1954}}{(1-e^2)^{3/2}},}
 which together with the Peters' equations \citep{Peters1964} allows us to compute the eccentricity when the binary enters the LIGO-Virgo-KAGRA frequency band ($f \simeq 10\text{ Hz}$). What we find is that, due to the large initial value of $a$, the eccentricity is radiated away before the GW emission becomes observable and thus the waveform appears nearly circular. For the above values of $M$, $r_h$, $a$ and $m_\BH$, we  find a merger that has circularised to $e_{10 Hz}\sim 10^{-2}$, which is currently not detectable \citep{Lower2018}, but it is predicted that third generation GW detectors should be able to find this population of BBH mergers, if it exists. Furthermore, if we assume that after each resonant intermediate state the eccentricity is sampled from a thermal distribution \citep{Heggie1975}, about $\sim 63\%$ ($\sim 0\%$) of captures that start with an initial semimajor axis of $a_h$ ($a_{ej}$) have a remaining $e_{10 Hz}<0.1$ at 10 Hz. The sampling of $e$ has a chance of producing mergers that form in the LIGO-Virgo-KAGRA band, $f > 10\text{ Hz}$, with a probability of $\sim 12\%$ ($\sim 46\%$). 

An alternative path to mergers via binary-binary interactions is the formation of stable triples. Stable triples with sufficiently high eccentricity can have large jumps in the eccentricity of the inner orbit (Lidov-Kozai cycles, per \citealt{Lidov1962, Kozai1962}). As above, extreme values of the eccentricity may lead to GW captures. Per the same argument, the involved semimajor axes may be sufficiently large that the eccentricity may be of the order of $e_{10 Hz}\sim 10^{-2}$ when the binary enters the LIGO-Virgo-KAGRA band. Modelling with  post-Newtonian terms is required to quantify the importance of mergers via these two channels.

The short timescale of the above processes implies that the merger happens at a location near the multiple-body interaction. This constitutes an explanation for the finding that the majority of mergers occur in-cluster -- as opposed to ejected binaries -- in \nbody models of low-mass clusters 
($M\lesssim10^5\,\msun$,  \citealt{Rastello2019}; \citealt{Banerjee2021b}; \citealt{Barber2023}). These models have $N\lesssim 1.5\times 10^5$, and for their initial mass function and metallicity $\mu\simeq0.04$ and $q\simeq 1/50$ such that $N_{BH}\lesssim100$, which is similar to the values considered in our two-mass \nbody models. As we explored different $q$ and $\mu$, our results can account for the different metallicities and other parameters in the more realistic models. The high fraction of in-cluster mergers found in these \nbody models are in contrast to fast models that assume a single active binary, such as \textsc{cBHBd} \citep{Antonini2023}, that predicts an in-cluster merger fraction of $\sim 40\%$. Although their assumption of a single, hard binary in the core at any given time is supported by our \nbody models, we here show that an important ingredient is missing in these fast models that would increase the  contribution of dynamically assembled BBH mergers in relatively low-mass star clusters.

Banerjee (priv. communication) indeed finds in-cluster mergers that form with a high $a$ ($\gtrsim 100\,\rm{AU}$) and extremely radial orbit, which nearly circularise at 10 Hz. These mergers may be a signature of binary-binary interactions at the hard-soft boundary. These binaries  contribute to the eccentricity distribution by increasing the expected rate of mergers at lower eccentricities ($e\gtrsim 10^{-2}$). Such mildly eccentric mergers could be disentangled from quasi-circular mergers in future GW detectors, such as the Einstein Telescope \citep{ETScienceCase} and Cosmic Explorer \citep{CEScienceCase}. In conclusion, we present an additional step towards a complete prediction for the rate of dynamically-formed BBH mergers that can be detected in current and upcoming  GW experiments. %


\section*{Acknowledgements}
The authors thank the referee Nathan Leigh for helpful comments and suggestions. The authors acknowledge financial support from the grants PRE2020-\allowbreak 091801, PID2021-\allowbreak 125485NB-\allowbreak C22, EUR2020-\allowbreak 112157, and CEX2019-\allowbreak 000918-\allowbreak M, funded by MCIN/\allowbreak AEI/\allowbreak 10.13039/\allowbreak 501100011033 (State Agency for Research of the Spanish Ministry of Science and Innovation), and SGR-2021-01069 grant (AGAUR).

\section*{Data availability}
The data underlying this article will be shared on reasonable request to the corresponding author.


\bibliographystyle{mnras}
\bibliography{bibliography, bibliography_books, bibliography_nonads, bibliography_lvk}{}


\bsp	
\label{lastpage}
\end{document}